\documentclass[a4paper]{article}
\usepackage{graphicx}

\title{MATHEMATICAL MODEL FOR HIT PHENOMENA AND ITS APPLICATION TO ANALYZE POPULARITY OF WEEKLY TV DRAMA}

\author{AKIRA ISHII, AKIKO KITAO, TSUKASA USUI $^{1}$, KOKI UCHIYAMA 	$^{2}	$ \\
\\
Department of Applied Mathematics and Physics, Tottori University, Koyama\\
Tottori-City, Tottori  680-8552, Japan\\
ishii@damp.tottori-u.ac.jp\\
{1} Data Co.Ltd., 1-16-16 Toranomon, Minato-ku\\
Tokyo 105-0001, Japan\\
{2} Hottolink, Kanda-nishikicho, Chiyoda-ku, \\
Tokyo 101-0054, Japan
}
\date{}
\begin{document}

\maketitle

\begin{abstract}
Mathematical model for hit phenomena presented by A Ishii et al in 2012 has been extended to analyze and predict a lot of hit subject using social network system. The equation for each individual consumers is assumed and the equation of social response to each hit subject is derived as stochastic process of statistical physics. The advertisement effect is included as external force and the communication effects are included as two-body and three-body interaction. The applications of this model are demonstrated for analyzing population of weekly TV drama. Including both the realtime view data and the playback view data, we found that the indirect communication correlate strongly to the TV viewing rate data for recent Japanese 20 TV drama. 
\end{abstract}


\section{Introduction}

Human interaction in real society can be considered in the sense of “many body” theory where each person can be treated as atoms or molecules in the ordinary many body theory of theoretical physics. Thus, even if we cannot use Hamiltonian because of lacking of energy conservation rules in social phenomena, we can use many body theory to the modeling of social phenomena in some sense. On the other hand, with the popularization of social network systems (SNS) like blogs, Twitter, Facebook, Google+, and other similar services around the world, interactions between accounts can be stocked as digital data. Though the SNS society is not the same as real society, we can assume that communication in the SNS society is very similar to that in real society. Thus, we can use the huge stock of digital data of human communication as observation data of real society \cite{Allsop,Kostka,Bakshy,Jansen}. Using this observed huge data (so -called "Big Data"), we can apply the method of statistical physics to social sciences. Since word-of-mouth (WOM) is very significant in marketing science \cite{Brown,Murray,Banerjee,Taylor}, such analysis and prediction of the digital WOM in the sense of statistical physics become very important today. 
Recently, we present a mathematical theory for hit phenomena where effect of advertisement and propagation of reputation and rumors due to human communications are included as the statistics physics of human dynamics \cite{Ishii2012a}. This theory has also been applied to the analysis of the local entertainment events in Japan successfully \cite{Ishii2012b}. Our model has been also applied to "general election" of a Japanese pop girls gourd AKB48 \cite{Ishii2013a}, music concert \cite{Kawahata2013a,Kawahata2013b} and even to a Kabuki player of 19th century\cite{Kawahata2013c}.

Our model was originally designed to predict how word-of-mouth communication spread over social networks or in the real society, applying it to conversations about movies in particular, which was a success. Moreover, we also found that when they overlapped their predictions with the actual revenue of the films, they were very similar. 

 In the model \cite{Ishii2012a}, the key factors to affect the mind of the persons in the society are three: advertisement or public announcement effects, the word-of-mouth (WOM) effects and the rumor effects. Recognizing that WOM communication, as well as advertising, has a profound effect on whether a person pays attention to the something or not, whether this is talking about it to friends (direct communication or WOM) or overhearing a conversation about it in a cafe (indirect communication or the rumor), we accounted for this in our calculations. The difference between our theory and the previously presented researches \cite{Elberse,Liu,Duan2008a,Duan2008b,Zhu,Goel,Karniouchina,Sinha,Pan,Asur,Ratkiewicz,Eliashberg,Bass1969,Bass1986,Dellarocas2004,Dellarocas2007} are discussed in ref.\cite{Ishii2012a}. 
 
We found that the effects of advertisements and WOM are included incompletely and the rumor effect is not included in the previous works \cite{Elberse,Liu,Duan2008a,Duan2008b,Zhu,Goel,Karniouchina,Sinha,Pan,Asur,Ratkiewicz,Eliashberg,Bass1969,Bass1986,Dellarocas2004,Dellarocas2007}. Therefore, from the point of view of statistical physics, we present in our previous paper a model to include these three effects: the advertisement or public announcement effect, the WOM effect, and the rumor effect. The previous model called "mathematical model for hit phenomena" has been applied to the motion picture business in the Japanese market, and we have compared our calculation with the reported revenue and observed number of blog postings for each film.

  However, in the recent our several works, we found that our theory can be applicable not only for box office but also other social entertainment like local events\cite{Ishii2012b}, animation drama on TV\cite{Ishii2013b},"general election" of a Japanese pop girl group AKB48\cite{Ishii2013a}, online music\cite{Ishii2012c}, play\cite{Kawahata2013d}, music concerts\cite{Kawahata2013a,Kawahata2013b}, Japanese stage actors\cite{Kawahata2013e}, Kabuki players of 19th century\cite{Kawahata2013c} and TV drama\cite{Ishii2014}. In these works, we have used an extended mathematical theory for hit phenomena for applying to general entertainments in societies. We can also applied the theory to analyze social phenomena. The theory has been applied to non-entertainment social phenomena; the scandal of the stimulus-triggered acquisition of pluripotency cell (known as STAP cell).
  \cite{Ishii-STAP2014a,Ishii-STAP2014b} In these works, we have used an extended mathematical theory for hit phenomena for applying to general entertainments in societies. 
  
  In this paper, the responses in social media are observed using the social media listening platform presented by Hottolink. Using the data set presented by M Data Co.Ltd monitors the exposure of each TV drama programs.

\section{Theory}

\subsection{Intention for individual person}

Based on the observation of posting on blog or twitter, we present a theory to explain and predict hit phenomena. First, instead of the number of potential persons who feel attention to the certain fact $N(t)$, we introduce here the integrated intention of individual persons, $J_i(t)$ defined as follows,

\begin{equation} 
\label{eq:eq3}
N(t) = \sum_i J_i(t)
\end{equation}

here the suffix $i$ corresponds to individual person who has attention to the event he/she concern. 

The daily intention is defined from $J_i(t)$ as follows, 
\begin{equation}
\frac{dJ_i(t)}{dt} = I_i(t)
\end{equation}

The number of integrated customers or incoming people to the event can be calculated using the intention as follows,

\begin{equation}
N(t) = \int_0^t \sum_i I_i(\tau) d\tau
\end{equation}

  Since the purchase intention of the individual customer increase due to both the advertisement and the communication with other persons, we construct a mathematical model for social  phenomena as the following equation. 

\begin{equation}
\frac{dI_i(t)}{dt} =  advertisement(t) + communication(t)
\end{equation}

\subsection{advertisement and communication}

Advertisement is the very important factor to increase intention of the customer in the market. Usually, the advertisement campaign is done at TV, newspaper and other media. We consider the advertisement effect as an external force to the equation of intention as follows,

\begin{equation}
\label{eq:eq12}
\frac{dI_i(t)}{dt} = \sum_{\xi} c_{\xi}A_{\xi}(t) + \sum_j d_{ij} I_j(t) * \sum_j \sum_k p_{ijk} I_j(t) I_k(t)
\end{equation}

where $A_{\xi}(t)$ is advertisement of each media like TV, newspaper, internet news sites, etc for each day in unit of counts or the exposure time, $c_{\xi}$ is the the factor of the effect of each advertisement and the suffix $\xi$ means the type of the media like TV, newspaper, internet news sites, etc. The factor $c_\xi$ are determined by using the Monte Carlo like method to adjust the observed data as we introduced in ref.\cite{Ishii2012a}. $d_{ij}$ is the factor for the direct communication and $p_{ijk}$ is the factor for the indirect communication\cite{Ishii2012a}. The suffix "$j$, $k$" are summed up for all friends for each person.  Because of the term of the indirect communication, this equation is a nonlinear equation.

\subsection{2 persons problem}

If we apply the equation (\ref{eq:eq12}) for 2 person problem, the equation can be written in the following form.

\begin{equation}
\label{eq:eq12a}
\frac{dI_1(t)}{dt} = \sum_{\xi} c_{\xi,1}A_{\xi,1}(t) +  d_{11} I_1(t) +  d_{12} I_2(t) 
\end{equation}

\begin{equation}
\label{eq:eq12b}
\frac{dI_2(t)}{dt} = \sum_{\xi} c_{\xi,2}A_{\xi,2}(t) +  d_{21} I_1(t) +  d_{22} I_2(t) 
\end{equation}

where the indirect communication term is not included because of only two person. If we neglect the effects of external mass media or environments, we obtain, 

\begin{equation}
\label{eq:eq12c}
\frac{dI_1(t)}{dt} =   d_{11} I_1(t) +  d_{12} I_2(t) 
\end{equation}

\begin{equation}
\label{eq:eq12d}
\frac{dI_2(t)}{dt} =   d_{21} I_1(t) +  d_{22} I_2(t) 
\end{equation}

These equations eqs.(\ref{eq:eq12c}), (\ref{eq:eq12d}) are equivalent to the equation of {\it "Love Affair"} introduced by Strogatz as the problem of love affair of Romeo and Juliet. \cite{Strogatz1,Strogatz2} Thus, it means that the Love Affair equation presented by Strogatz is included in our mathematical model for social phenomena. 

If we add the effect of external field like mass media to the above equations(\ref{eq:eq12c}), (\ref{eq:eq12d}), we can write the equations a follows. 

\begin{equation}
\label{eq:eq12e}
\frac{dI_1(t)}{dt} = c_{1}A_{1}(t) +  d_{11} I_1(t) +  d_{12} I_2(t) 
\end{equation}

\begin{equation}
\label{eq:eq12f}
\frac{dI_2(t)}{dt} = c_{2}A_{2}(t) +  d_{21} I_1(t) +  d_{22} I_2(t) 
\end{equation}

The above is equivalent to the work of Wauer et.al. \cite{Wauer} where the external effects are added in the Love Affair equations of Strogatz. Thus, our model include also the work of Wauer et.al.\cite{Wauer} for modified Love Affair equation. In that sense, our mathematical model for social phenomena is the extension of the Love Affair equation by Strogatz to many person problem including indirect communication effects.

\subsection{Mean field approximation}

To solve the equation (\ref{eq:eq12}), we introduce here the mean field approximation for simplicity. Namely, we assume that the intentions of every persons are similar so that we can introduce the averaged value of the individual intention. 

\begin{equation}
I = \frac{1}{N_p} \sum_j I_j(t)
\end{equation}

where introducing the number of potential customers $N_p$.  We obtain the direct communication term from the person who do not watch the movie as follows,

\begin{equation}
\sum_j d_{ij} I_j(t) = d \sum_j I_j(t) = N_p d I(t) = D I(t)
\end{equation}

where $D = N_p d$ is the averaged factor of the direct communication. 

Similarly, we obtain the indirect communication, 

\begin{eqnarray}
\sum_j \sum_k p_{ijk} I_j(t) I_k(t) = p \sum_j \sum_k I_j(t) I_k(t) = N_p^2 p I^2(t) \nonumber  \\
= P I^2(t)
\end{eqnarray}

where $P = N_p p$ is the averaged factor of the indirect communication.

Substituting the above, we obtain the following equation as the equation of intention, 

\begin{equation}
\label{eq:eq13}
\frac{dI(t)}{dt} = \sum_{\xi} c_{\xi}A_{\xi}(t) + D I(t) + P I^2(t). 
\end{equation}

The above equation can be also derived as the following way. If we consider the unknown function for the effect of human communication as $F(I(t))$ to write

\begin{equation}
\frac{dI(t)}{dt}= \sum_{\xi} c_{\xi}A_{\xi}(t) + F(I_0 + I(t)),
\end{equation}

we can expand the function $F$ by $I(t)$ as follows, 

\begin{equation}
\label{eq:eq14}
\frac{dI(t)}{dt}= \sum_{\xi} c_{\xi}A_{\xi}(t) + F(I_0) + \frac{dF}{dI} I(t) 
+ \frac{1}{2} \frac{d^2F}{dI^2} I^2(t) +  \frac{1}{3!} \frac{d^3F}{dI^3} I^3(t) + 
\cdots
\end{equation}

Thus, we can recognize that the equation (\ref{eq:eq13}) corresponds to the  equation (\ref{eq:eq14}) of the second-order.

The equations  (\ref{eq:eq13}) is based on the equation we presented in the previous work for the motion picture entertainment market \cite{Ishii2012a} where the equation is derived using the stochastic processes. Thus, the equation  is the nonlinear differential equation. However, since the handling data is daily, the time difference is one day, we can solve the equation numerically as a difference equation.

\subsection{Determination of parameters}

 For the purpose of the reliability, we introduce here the so-called “R-factor” (reliable factor) well-known in the field of the low energy electron diffraction (LEED) experiment \cite{Pendry}. In the LEED experiment, the experimentally observed curve of current vs. voltage is compared with the corresponding theoretical curve using the R-factor. 
 For our purpose, we define the R-factor for our purpose as follows,

\begin{equation}
\label{eq:Rfactor}
R = \frac{\sum_i (f(i)-g(i))^2}{\sum_i (f(i)^2 - g(i)^2)}
\end{equation}

where the functions $f(i)$ and $g(i)$ correspond to the calculated $I(t)$ and the observed number of posting to blog or Twitter. The smaller R, the function $f$ and $g$ show that better matches. Thus, we use the random number to search the best parameter set which gives us the minimum R. This random number technique is similar to the Metropolis method.\cite{Metropolis}   This method has been introduced in ref.\cite{Ishii2012a} We use this R-factor as guide to get best adjustment of our parameters for each calculation in this paper.

Actually, the parameters $c_{\xi}$、$D$ and $P$ in eq.(\ref{eq:eq13}) can be considered as a function of time, because the attention of people can be changed as a function of time. However, if  we introduce the functions $c_{\xi}(t)$、$D(t)$ and $P(t)$, we can adjust any phenomena by adjusting these functions. Thus, we try to keep the parameters $c_{\xi}$、$D$ and $P$ to be constant value in order to examine that the equation (\ref{eq:eq13}) can be apply to any social phenomena or not.

\subsection{Time difference in the actual caculation}

In actual calculation, the equation of the mathematical theory of hit phenomena is not solved as differential equation but a difference equation, 

\begin{equation}
\label{eq:eq15}
\Delta I(t) = [\sum_{\xi} c_{\xi}A_{\xi}(t) + D I(t) + P I^2(t) ] \Delta t. 
\end{equation}

In previous papers using the mathematical theory of hit phenomena for many phenomena on entertainment or social incident\cite{Ishii2012a,Ishii2012b,Ishii2013a,Kawahata2013a,Kawahata2013b,Kawahata2013c,Ishii2013b,Ishii2012c,Kawahata2013d,Ishii2014,Ishii-STAP2014a,Ishii-STAP2014b}, the time difference $\Delta t$ is chosen to be one day, because blog is considered to be posted only one post per one day at night. It is very good approximation for blog analysis. For the case of analysis of Kabuki player of 19th century in Japan, the time difference $\Delta t$ is chosen to be one month, because the {\it big data} of 19th century, the {\it ukiyoe picuture} can be followed only per month.\cite{Kawahata2013c}  

In the case of Twitter, many people post their tweet every time, so that the time difference $\Delta t$ in the mathematical theory of hit phenomena can be chosen to be far shorter. In this paper, we chose $\Delta t$ to be one hour. In principle, we can chosen $Delta t$ to be much shorter, one second, for example.

\section{Results}

\subsection{Examples of application for Box Office of Movie}

First, we show an example of our calculation for the box office of movie, "Dark Night Rises" using the theory of ref.\cite{Ishii2012a}. In the figure, the histogram means the exposure time of the movie on TV in Japan and we use the data as input data of $A(t)$, where we use only single media for advertisement. We compare our calculation with the observed daily Twitter posting on the movie. The parameters C, D and P are determined by using the Monte Carlo way\cite{Ishii2012a}.

\begin{figure}
\centering
\includegraphics[height=6.5cm, bb=0 0 720 540]{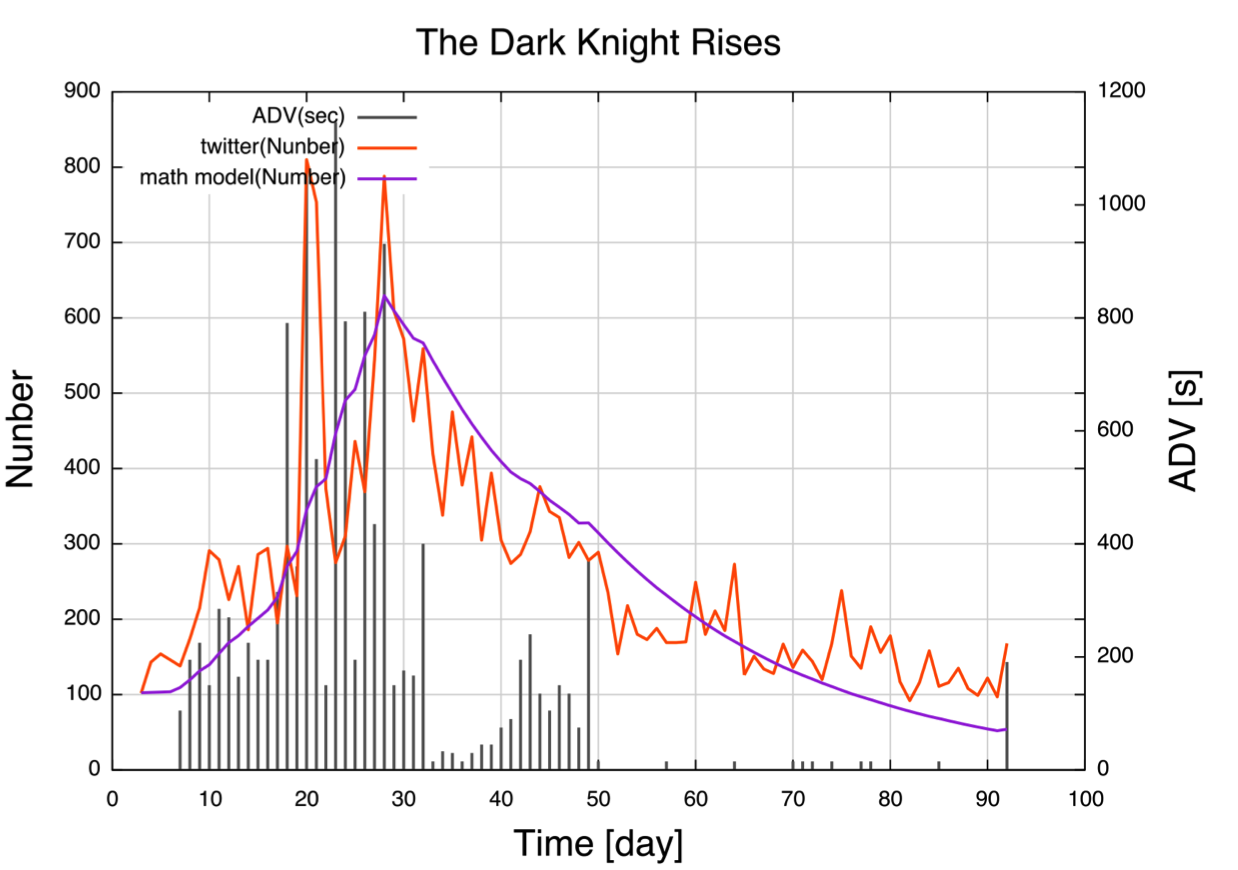}
\caption{Observed twitter posting, advertisement and our calculation for the Dark Night Rises.}
\label{fig:Dark}
\end{figure}

\subsection{Digital TV log data}

For analyzing the TV viewing rate, the most problem is that Japanese TV viewing rate issued by Video Research Ltd.  is calculated from the television volume looking only in real time. However, in the actual situation, many people watch drama using the playback of the recorded drama. Since people post as blog or Twitter after watching each drama, we should know the recording situation for TV drama of each person in society. Such informations had been impossible to obtain before the digital TV system began. 

Nowadays, the number of digital television sets (digital TV) have become popular in the Japanese market. Using these digital TV, the TV makers can save log data of every operations of the TV controller of every digital TV sets as all records. Using these digital TV log data, we can collect the information of the recording and the playback of each drama on every digital TV sets.  For this study, we can use the digital TV data required in our analysis of the drama under the support of the Toshiba corporation. Because of the contract between us and Toshiba corporation, the exact number of real time view and recorded view for each drama cannot show in the paper, though we use such data.

\begin{figure}
\centering
\includegraphics[height=7cm, bb=0 0 750 550]{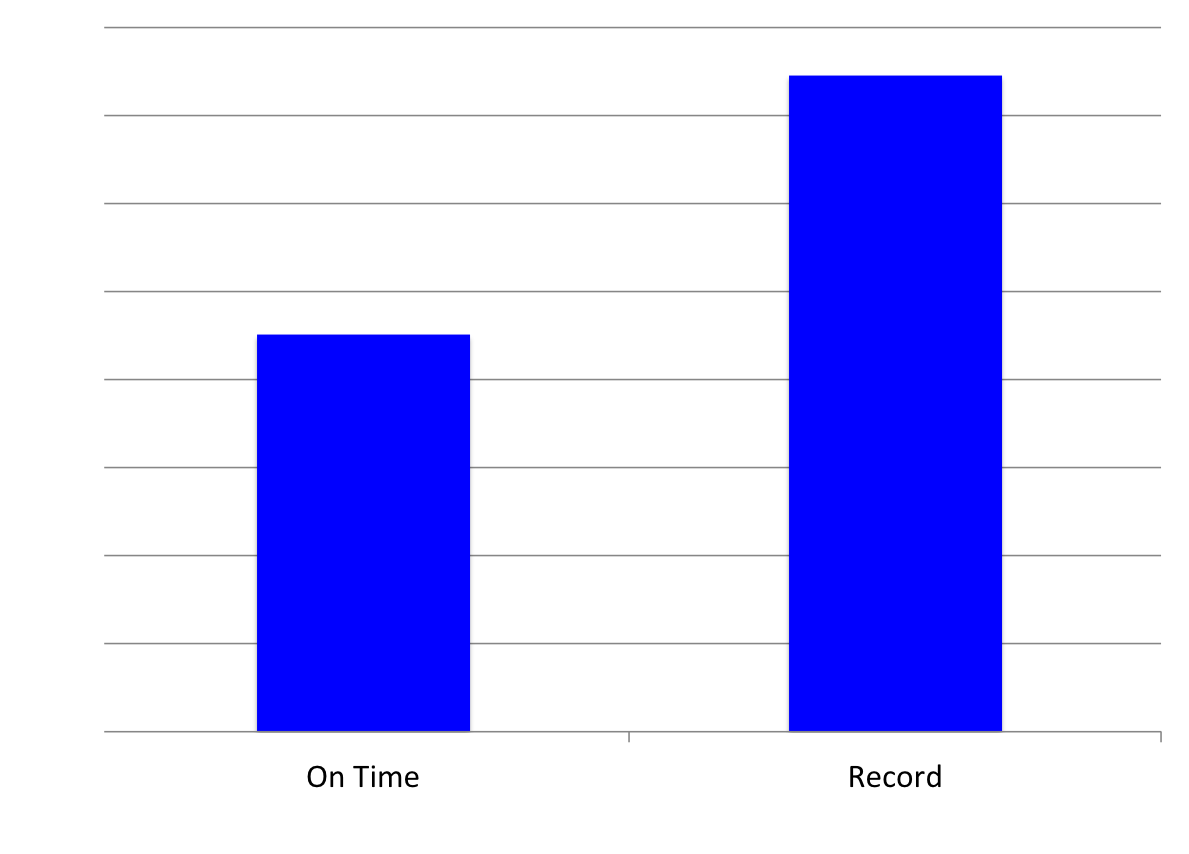}
\caption{The averaged counts of the realtime viewing and the recording of each drama we check in this study using the Toshiba digital TV log system.}
\label{fig:figA}
\end{figure}

In fig.\ref{fig:figA}, we show the averaged counts of the realtime viewing and the recording of each drama we check in this study. As shown in the figure, the counts of the recording is far more than the realtime viewing. 

\begin{figure}
\centering
\includegraphics[height=7cm, bb=0 0 650 500]{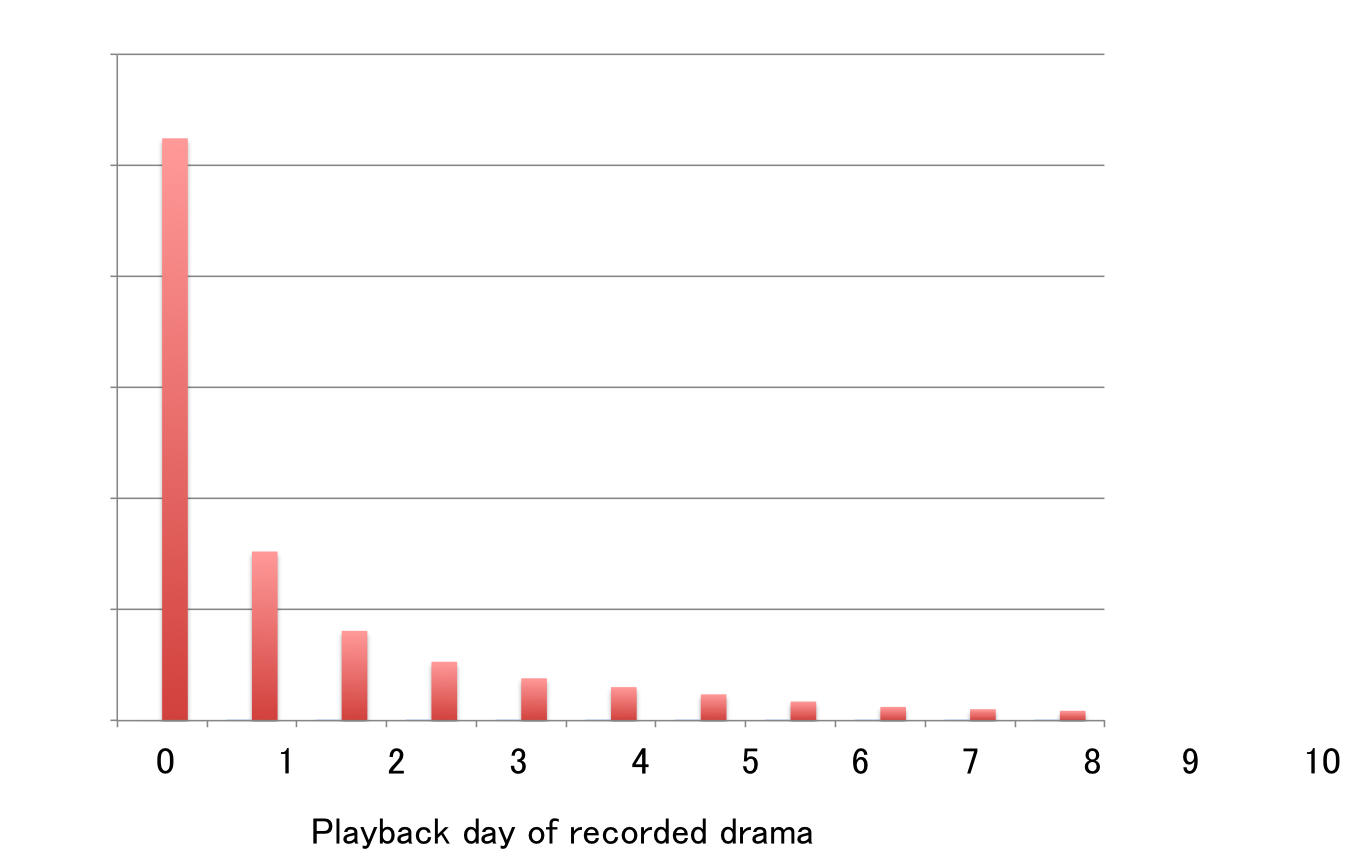}
\caption{The averaged playback data for all drama we analyzed in this study}
\label{fig:figB}
\end{figure}

In fig.\ref{fig:figB}, we show the actual playback data; the averaged playback data for all drama we analyzed in this study. Roughly, about half of the recorded data are played in the same day.  The remaining half of the play is distributed over 1 month. We can also checked that a few recorded data are not played. 

\begin{figure}
\centering
\includegraphics[height=7cm, bb=0 0 750 550]{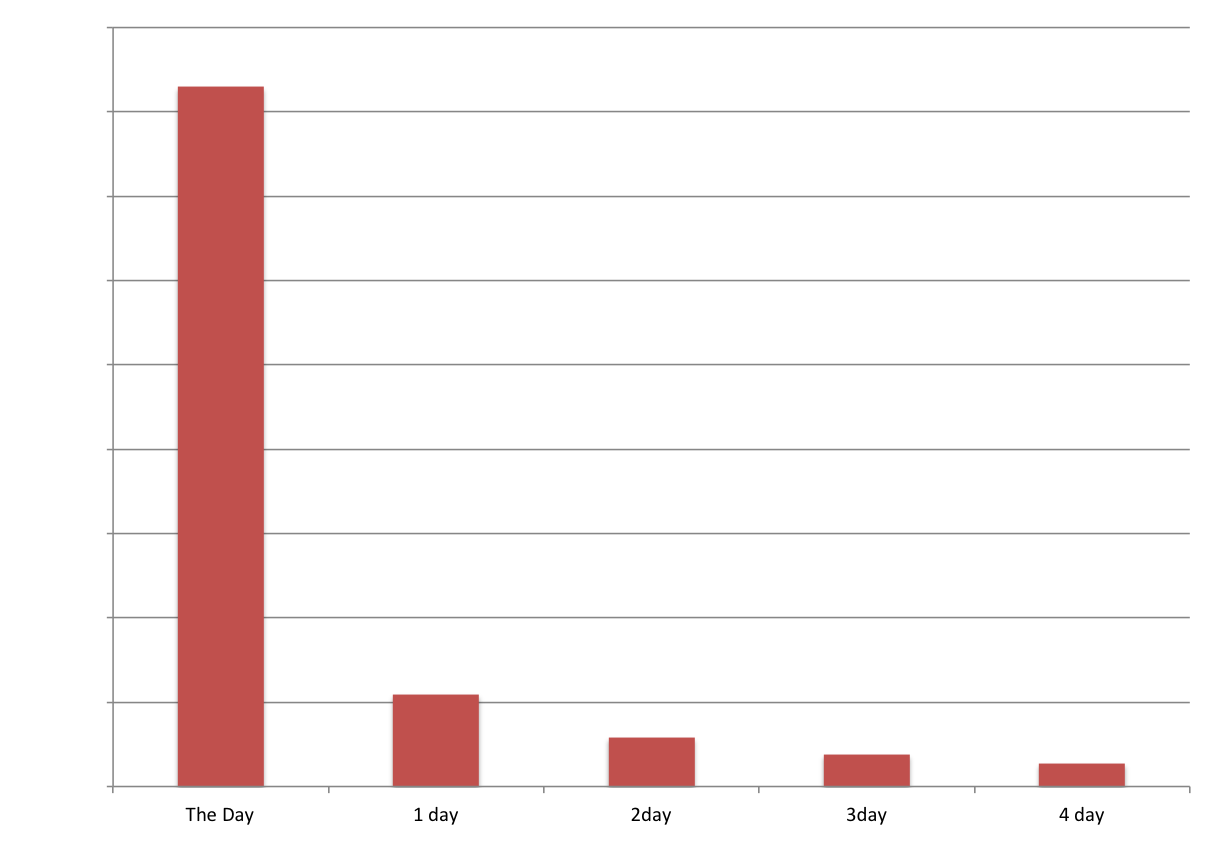}
\caption{The actual viewing data including both the realtime viewing and the playback viewing}
\label{fig:figC}
\end{figure}

In fig.\ref{fig:figC}, we show the actual viewing data including both the realtime viewing and the playback viewing. Actually, we have these data for every term of every drama we analyzed here separately. We use this data for $A(t)$ of (\ref{eq:eq13}) as the viewing data of each drama.

\begin{figure}
\centering
\includegraphics[height=7cm, bb=0 0 550 400]{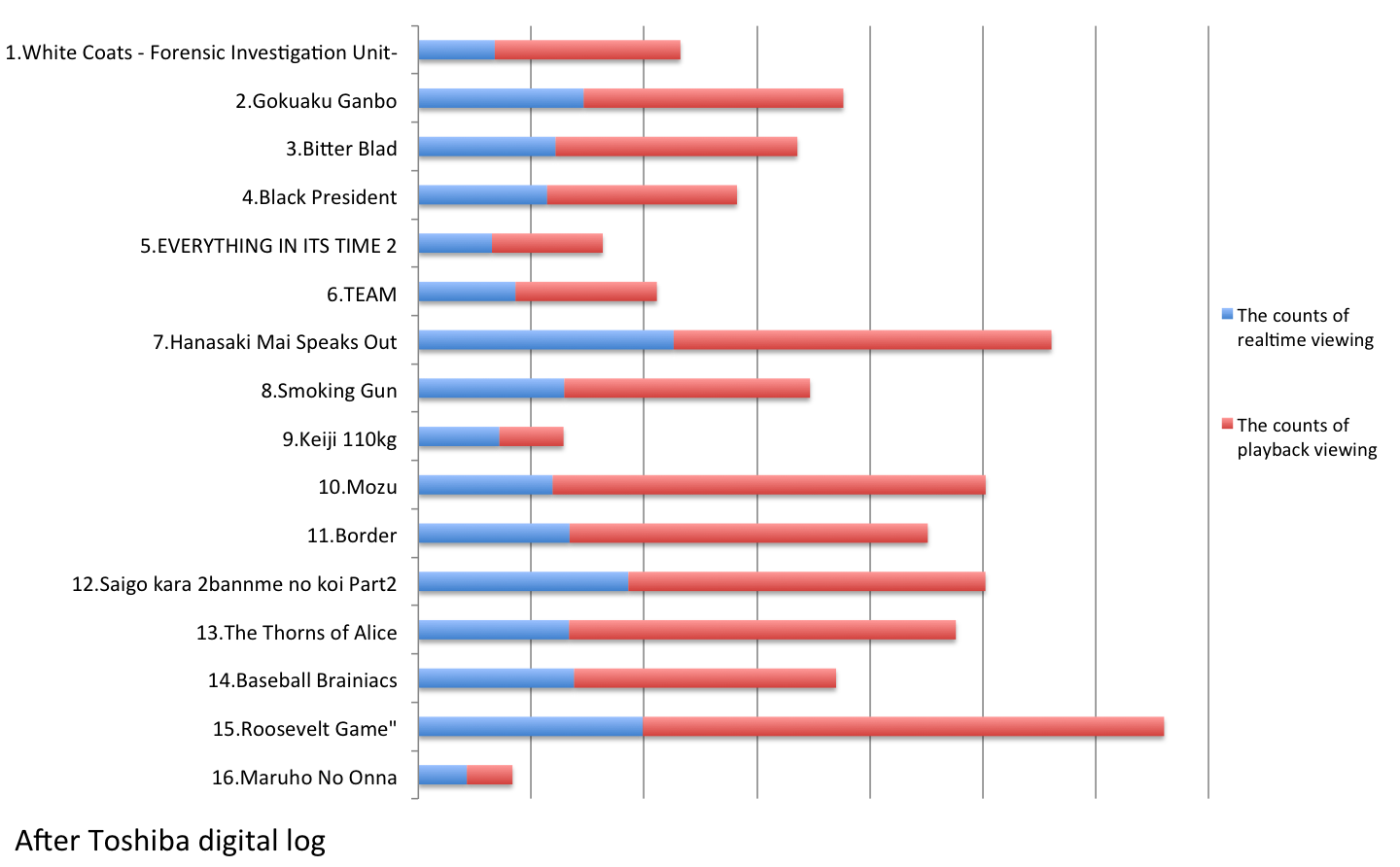}
\caption{The counts of the realtime viewing and the playback viewing from recorded data. The blue means the realtime viewing and the red means the playback viewing for 16 drama broadcasted in 2014 summer: drama 1 to 16 are "White Coats - Forensic Investigation Unit -", "Gokuaku Ganbo", "Bitter Blad", "Black President", "EVERYTHING IN ITS TIME 2", "TEAM", "Hanasaki Mai Speaks Out", "Smoking Gun", "Keiji 110kg", "Mozu", "Border", "Saigo kara 2bannme no koi Part2", "The Thorns of Alice", "Baseball Brainiacs", "Roosevelt Game", "Maruho No Onna", respectively. }
\label{fig:figD}
\end{figure}

In fig.\ref{fig:figD}, we show the counts of the realtime viewing and the playback viewing from recorded data. As we can see in the figure, the rate of the playback viewing is often more than the realtime viewing. For the drama 7 and the drama 15, the drama 7 has much realtime viewing, but the drama 15 is more if we count both the realtime viewing and the playback viewing. Thus, we can consider that the data of playback viewing is significant to analyze the actual popularity of each drama. 

\subsection{Blog and Twitter}

We can use social networks to analyze the popularity of TV drama. The frequently used social networks are Facebook, Twitter and blog. Unfortunately, we cannot obtain the data from Facebook, because most of all Japanese account holder of Facebook write their notes, diaries or comments only for their own friend. Thus, we cannot access most of all text in Facebook. Therefore, we usually use Twitter or blog for analysis using the mathematical theory for hit phenomena. 

\begin{figure}
\centering
\includegraphics[height=7cm, bb=0 0 700 500]{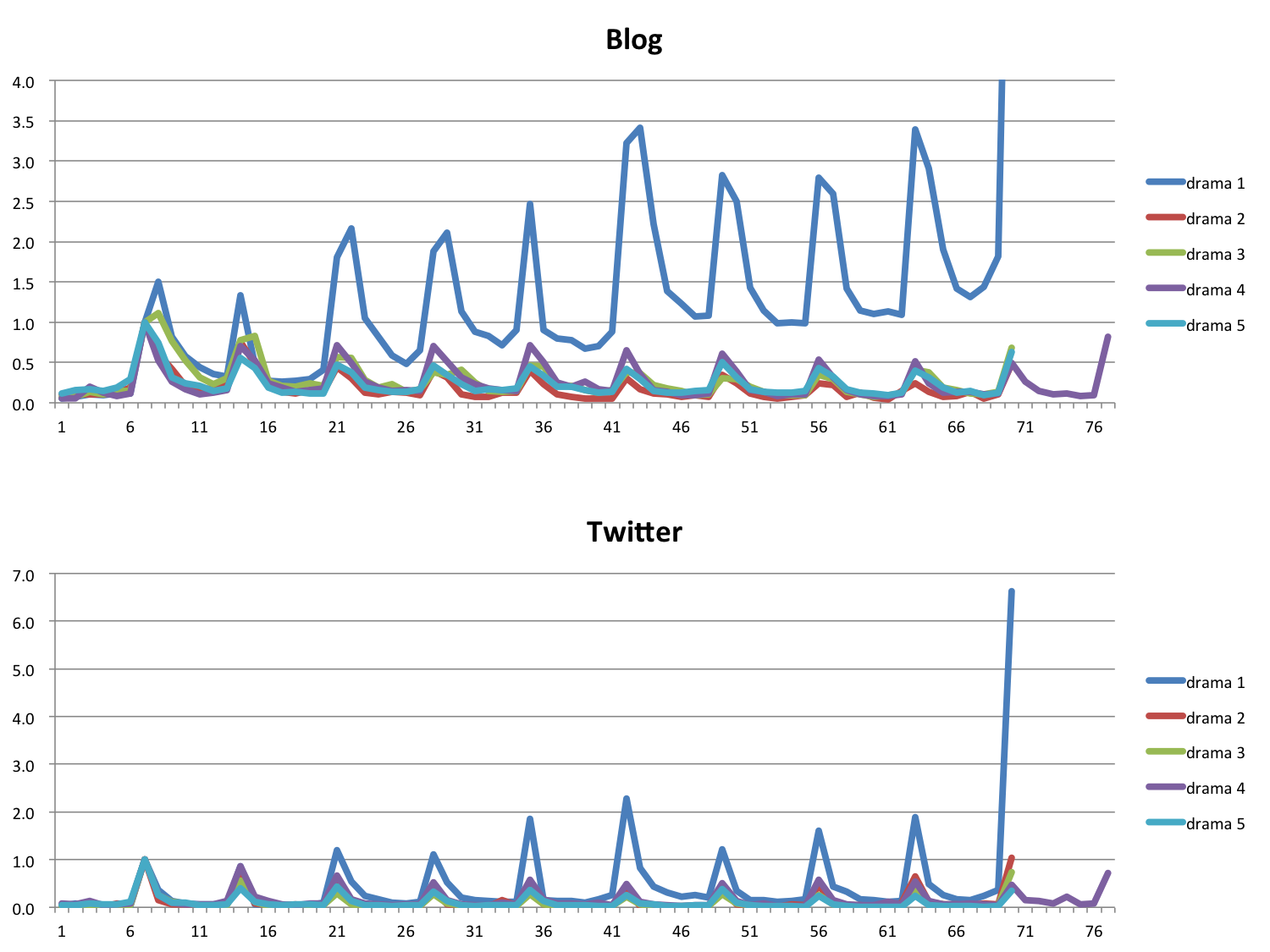}
\caption{The difference of the observed counts of posting for blogs and Twitter for 5 weekly drama we analyze in this paper. }
\label{fig:figE}
\end{figure}

In  fig.\ref{fig:figE}, we show the difference of the observed counts of posting for blogs and Twitter for 5 weekly drama we analyze in this paper.  For every drama, there are peaks of the posting to blogs or Twitter at the day of the broadcasting of the drama itself. The 5 drama we analyze are follows;

\begin{enumerate}
\item drama 1  {\it Hanzawa Naoki}, Business story
\item drama 2  {\it S The Last Policeman}, Police story
\item drama 3  {\it android -A.I. knows LOVE?}, Science Fiction
\item drama 4  {\it Un chocolatier de l'amour perdu}, Love story
\item drama 5  {\it Pintokona}, Love story
\end{enumerate}

Among the 5, the drama 1 {\it Hanzawa Naoki} is very success drama in Japan and it gets very high score of TV view rating in this couple of years. As we found in the figure,  especially for blog posting, the drama 1 {\it Hanzawa Naoki} has very high counts. For the drama 1 {\it Hanzawa Naoki}, we found a lot of posting even at the remain 6 days. However, for Twitter posting, even for many positing to the drama 1 {\it Hanzawa Naoki}, the high counts is only one day, the broadcasting day of the drama and we found very few posting in the remain 6 days. 

Therefore, we found that, for the analysis of the daily posting using the equation (\ref{eq:eq13}), the blog data is better than the Twitter data, because the time-dependent variation of postings are significant for this analysis.

\subsection{Effect of Playback viewing}

\begin{figure}
\centering
\includegraphics[height=7cm, bb=0 0 550 400]{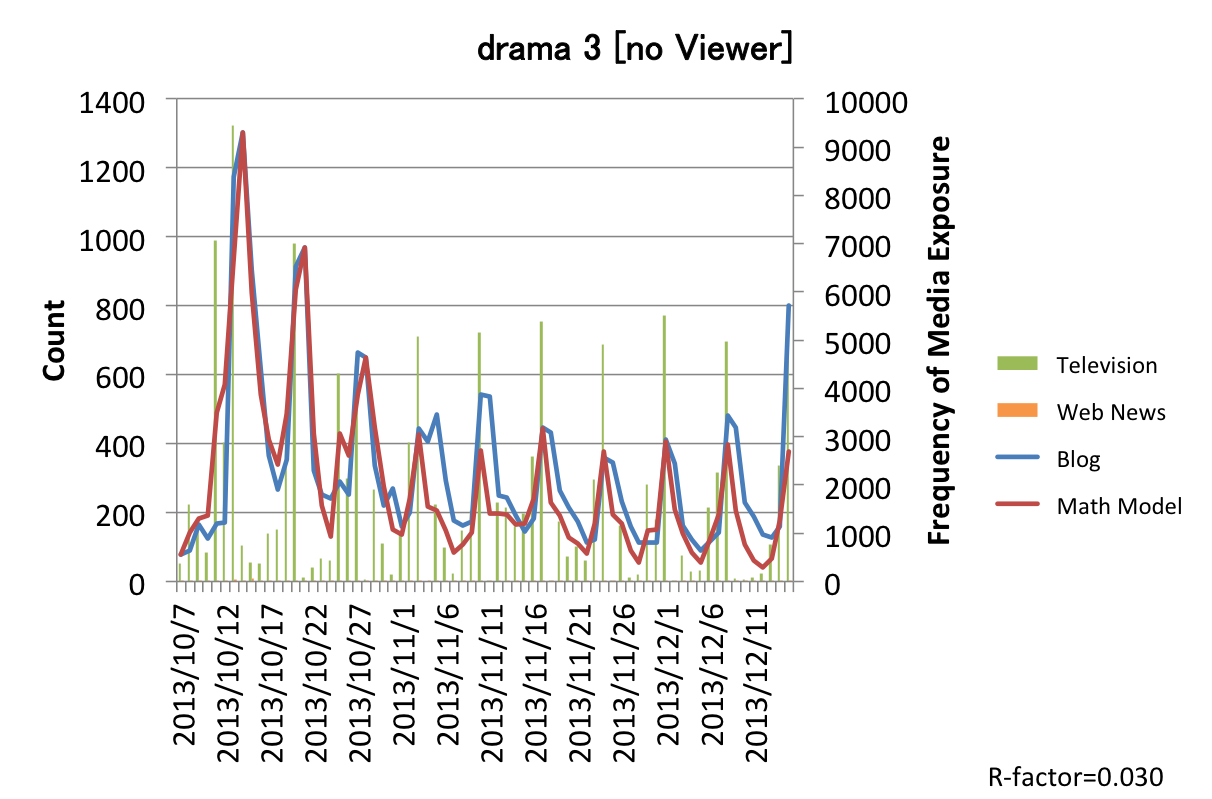}
\caption{The best calculated results for the drama   {\it android -A.I. knows LOVE? } without the playback viewing data. R-factor is 0.030}
\label{fig:figF}
\end{figure}

For the purpose of the reliability to determine parameters C, D and P of  eq(\ref{eq:eq13}), we introduce the so-called “R-factor” (reliable factor) in eq(\ref{eq:Rfactor}). Using the R-factor, we check the importance of the playback viewing effect. In  fig.\ref{fig:figF}, we show the best calculated results for the drama   {\it android -A.I. knows LOVE? } with and without the playback viewing data. For the better fitting, the R-factor is smaller. 

\begin{figure}
\centering
\includegraphics[height=7cm, bb=0 0 550 400]{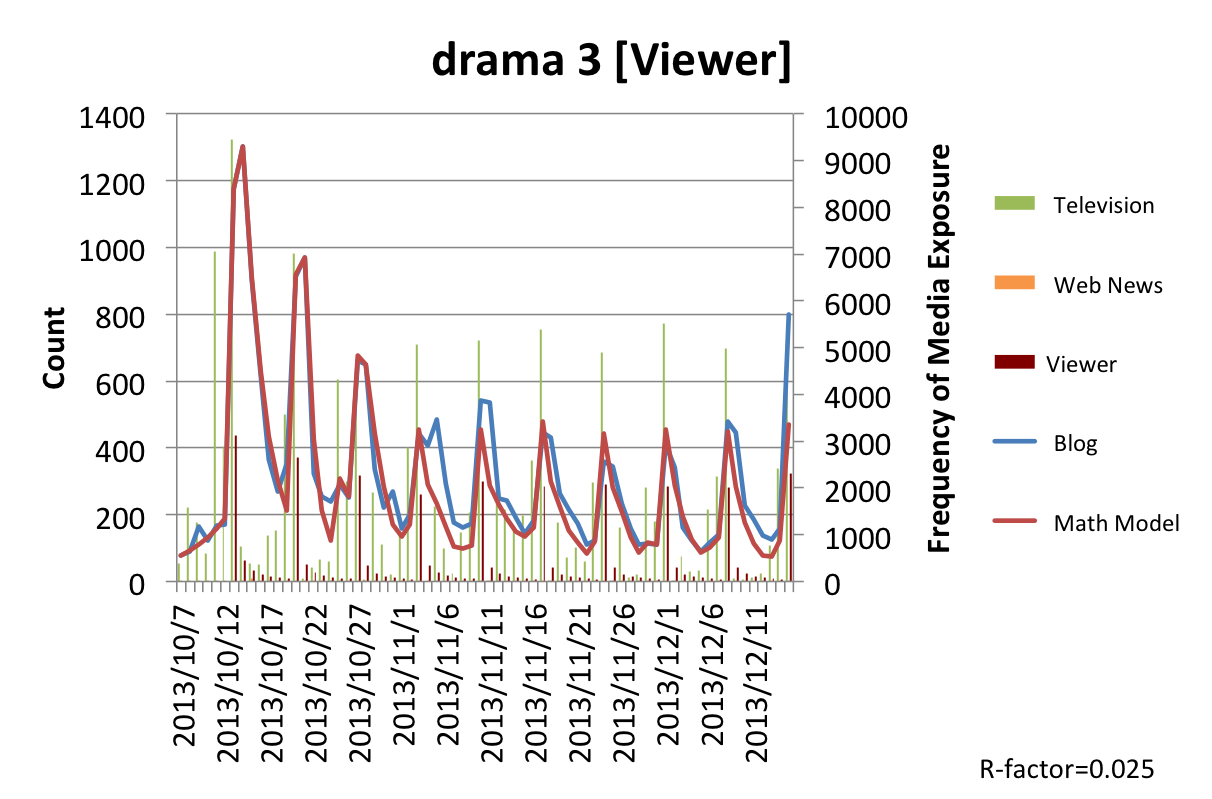}
\caption{The best calculated results for the drama   {\it android -A.I. knows LOVE? } with the playback viewing data. R-factor is 0.025.}
\label{fig:figG}
\end{figure}

As we can see in the figure captions, we obtain the smaller R-factor for the calculation including the playback viewing data. Thus, we found that the playback viewing effect is very important to analyze the popularity of TV drama. Moreover, it means that many people posts blogs after they watch each drama program independent of the viewing of other people.

\subsection{Correlation between the TV viewing rate and the parameters of this model}

\begin{figure}
\centering
\includegraphics[height=7cm, bb=0 0 550 400]{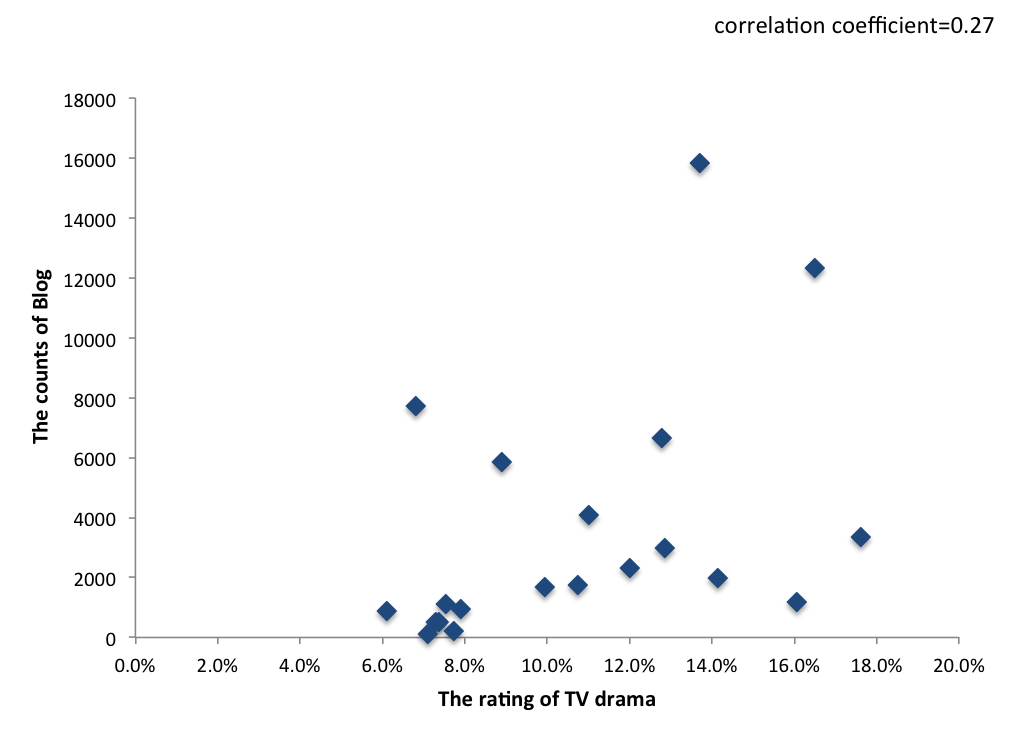}
\caption{The correlation between the TV viewing rate and the total number of blog posting for 20 drama of Japanese market in 2013 and 2014. The TV viewing rate we show in the figure is the values issued from the Video Research. The correlation coefficient is 0.27. }
\label{fig:figL}
\end{figure}

In fig.\ref{fig:figL}, we show the  correlation between the TV viewing rate and the total number of blog posting for 20 drama of Japanese market in 2013 and 2014. The TV viewing rate we show in the figure is the values issued from the Video Research. The correlation coefficient in this case is 0.27. We found similar results also for Twitter posting. Thus, as many people suppose, the total numbers of posting of blog and Twitter for TV drama have very  low correlation with the TV viewing rate. 

\begin{figure}
\centering
\includegraphics[height=7cm, bb=0 0 550 400]{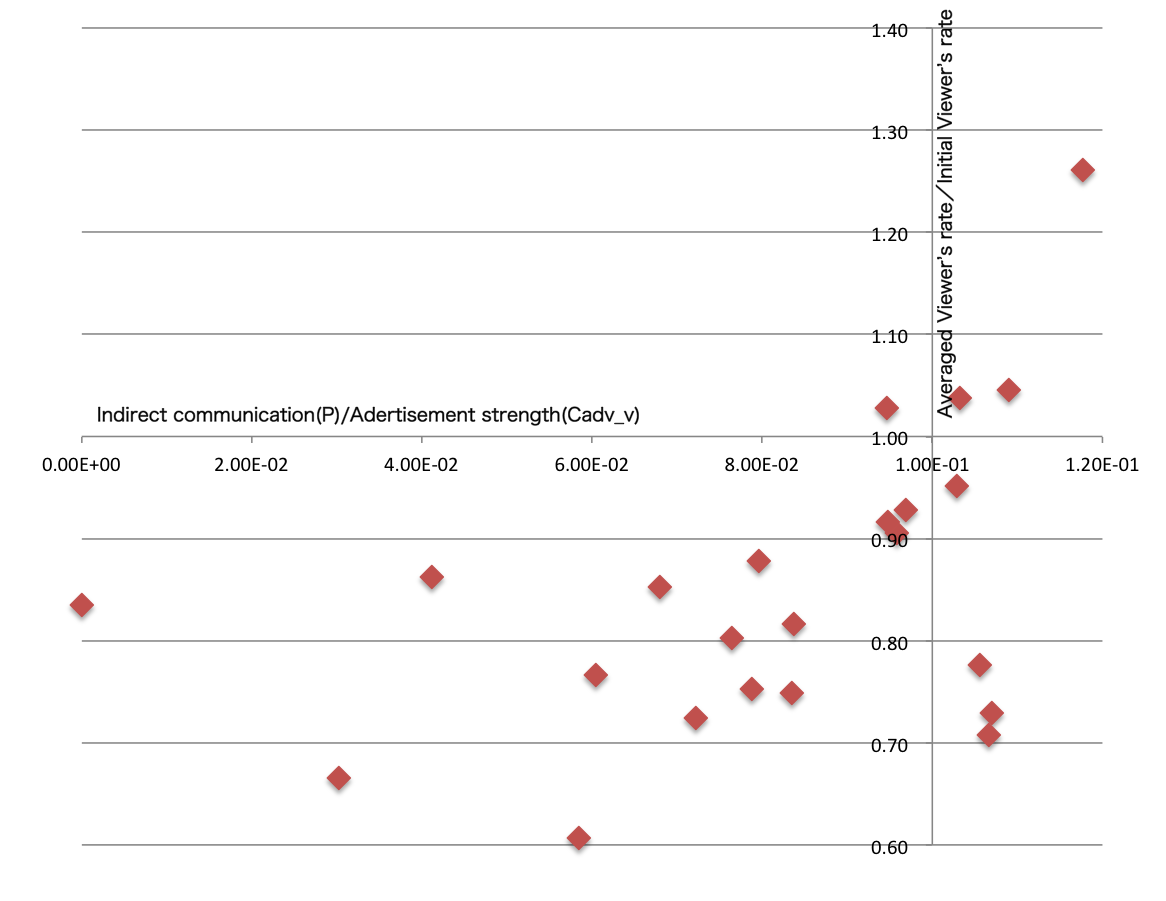}
\caption{The correlation between the TV viewing rate and the indirect communication factor for 20 drama of Japanese market in 2013 and 2014. The TV viewing rate we show in the figure is the averaged TV viewing rate of each drama divided by the viewing rate of the first term of each drama. The indirect communication strength here is the indirect communication strength P decided by the effective advertisement strength C for the drama viewing both for the realtime and the playback. The correlation coefficient is 0.78. }
\label{fig:figH}
\end{figure}

On the contrary, if we use the indirect communication strength instead of the total number of positing, we have very high correlation.  In  fig.\ref{fig:figH}, we show the correlation between the TV viewing rate and the indirect communication factor for 20 drama of Japanese market in 2013 and 2014. The TV viewing rate we show in the figure is the averaged TV viewing rate of each drama divided by the viewing rate of the first term of each drama. It means the increase or decrease of the viewing rate during the broadcasting period of each drama. The indirect communication strength we show in the figure is the indirect communication strength P decided by the effective advertisement strength C for the drama viewing both for the realtime and the playback. 

\begin{figure}
\centering
\includegraphics[height=7cm, bb=0 0 850 600]{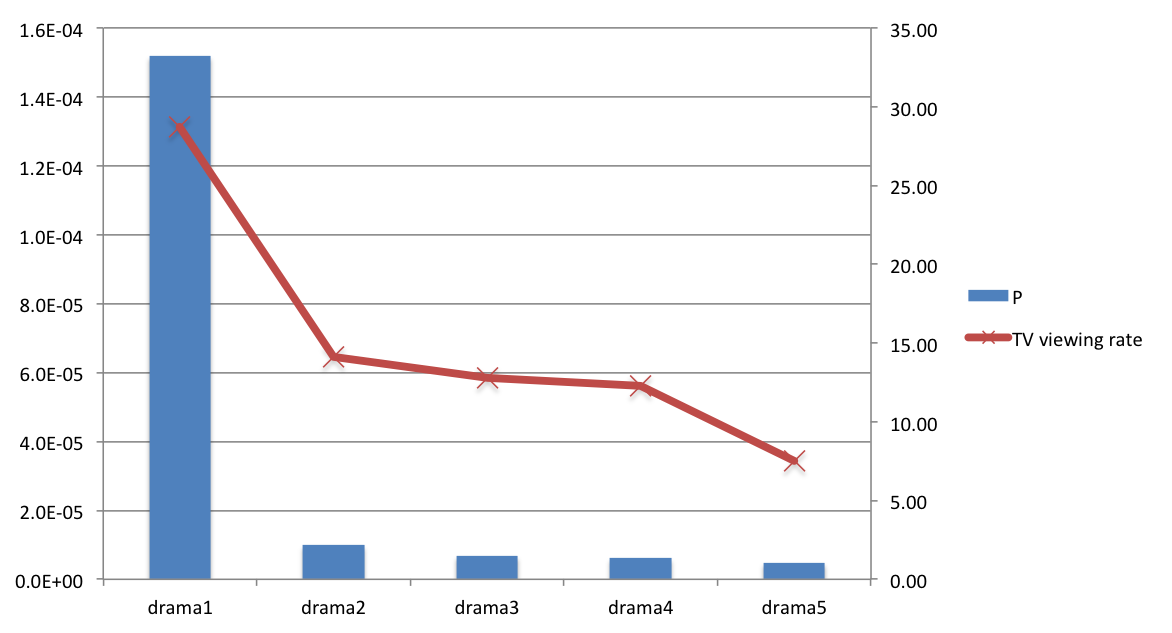}
\caption{The comparison between the TV viewing rate and the indirect communication for 5 drama}
\label{fig:figJ}
\end{figure}

\begin{figure}
\centering
\includegraphics[height=7cm, bb=0 0 850 600]{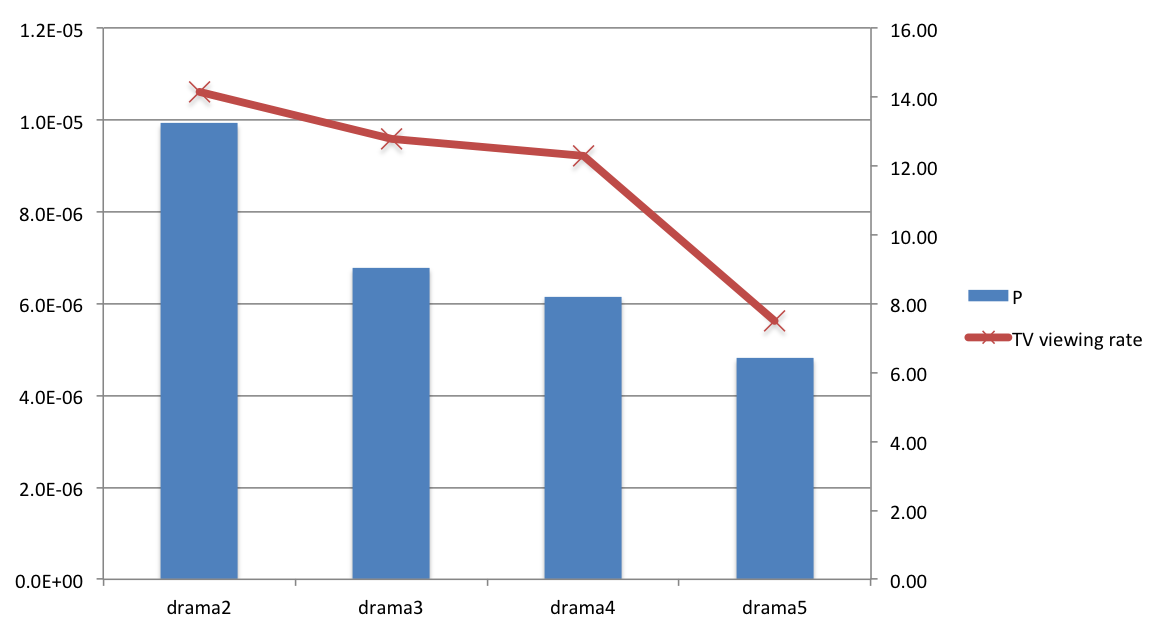}
\caption{The comparison between the TV viewing rate and the indirect communication for 4 drama}
\label{fig:figK}
\end{figure}

In the figure, we found the positive correlation between the viewing rate and the indirect communication strength. The correlation coefficient in this case is 0.78 where one drama, "Maruho no Onna" no,16 in fig.\ref{fig:figD} is excluded in the calculation of the correlation because it is the drama of very minor TV station.  In  fig.\ref{fig:figJ} and fig.\ref{fig:figK}, we show the comparison between the TV viewing rate and the indirect communication for 5 drama. We can also confirm that the TV viewing rate and the indirect communication for each drama is very similar. 

\section{Discussion}

Mathematical model for hit phenomena  \cite{Ishii2012a} is not only for the analysis and prediction of the box office business having clear open day of each movie but also for a variety of entertainments and social evidences having no clear open day or closed day. Even for periodic case, we can apply the model: weekly TV drama are the case of periodic having period of one week. In this paper, we use the mathematical model for hit phenomena to analyze weekly TV drama. 

The popularity of drama is related to the population of blog posting or twitter posting in some sense, because people post frequently their comments for watched drama.  However ,the discussion of the relation between the popularity of each drama and the posting of blog or tweets for each drama is not simple. For the evaluation of the popularity of drama, we should include the playback viewing of each drama using recorded media. Moreover, the total number of blog posting (or tweet posting) has almost no relation to the TV viewing ratio as we saw in fig.\ref{fig:figL} that we found the  correlation between the TV viewing rate and the total number of blog posting for 20 drama of Japanese market in 2013 and 2014 has the correlation coefficient of only 0.27. 

In this work, using the dataset presented by Toshiba co. ltd for digital TV log, we can include the effect of the playback viewing exactly. We found that the weight of the playback viewing is almost similar to the realtime viewing. We can also include the delay of the viewing from recorded files exactly from the real data of Toshiba. We found that the inclusion of the playback viewing is significant because we obtain better R-factor for the case where both the realtime and the playback viewing are included in the calculation of the mode.

In fig.\ref{fig:figH}, we found the very high correlation between the TV viewing rate and the strength of the indirect communication strength obtained by our mathematical theory for hit phenomena: the correlation coefficient is 0.78. The indirect communication strength P decided by the effective advertisement strength C for the drama viewing both for the realtime and the playback we used in fig.\ref{fig:figH} is the population of each drama per a certain advertisement.  Such high correlation is also confirmed in  fig.\ref{fig:figJ} and fig.\ref{fig:figK}. In the calculation, we include both the realtime viewing and the playback viewing from recorded drama. Thus, we can consider that the indirect communication is very significant to explain the viewing of drama by people in the society.  

Therefore, from this work, we can consider that the direct communication effect in the mathematical model for hit phenomena corresponds to the evaluation of drama quality by persons who watched drama. On the other hand,  the indirect communication effect in the model  corresponds to diffuse attractiveness of each drama to persons who has not yet watched the corresponding drama. 

Thus, to get very high population of drama, the two factors are significant; the advertisement before the start of the drama and the indirect communication strength. The  advertisement before the start of the drama is important to have high initial attention at the beginning of the drama. The indirect communication strength is significant to increase person who watch drama. Certainly, how to get high indirect communication strength for each drama is not problems for social physics but the problem for each drama director and actor/actress.

\section{Conclusion}

Mathematical model for hit phenomena is extended to analyze the popularity of weekly TV drama. The equation for each individual consumers is assumed and the equation of social response to each hit subject is derived as stochastic process of statistical physics. The advertisement effect is included as external force and the communication effects are included as two-body and three-body interaction. Using the model. we analyze the popularity of drama in Japanese market and we include both the realtime viewing and the playback viewing using the digital TV log data of Toshiba.  We found the high correlation between the TV viewing rate and the strength of the indirect communication strength obtained from the analysis of daily blog posting of each drama using our model.

\section*{Acknowledgments}

For this study, we can use the digital TV data required in our analysis of the drama under the support of the Toshiba corporation.

\end{document}